\documentclass[prf,superscriptaddress,showpacs,longbibliography,aps]{revtex4-1}

\usepackage[final]{graphicx}
\usepackage{amsmath}
\usepackage{verbatim}
\usepackage{color}
\usepackage{amssymb}
\usepackage{ulem}
\usepackage{natbib}
\usepackage{textcomp}
\usepackage[margins]{trackchanges}
\renewcommand{\vec}[1]{\mathbf{#1}}

\begin{document}

\title{Structure induced vortices control anomalous dispersion in porous media}

\author{Ankur Deep Bordoloi}
\affiliation{Institute of Earth Sciences, University of Lausanne, Lausanne 1015, Switzerland}

\author{David Scheidweiler}
\affiliation{Institute of Earth Sciences, University of Lausanne, Lausanne 1015, Switzerland}

\author{Marco Dentz}
\affiliation{Spanish National Research Council (IDAEA-CSIC), Barcelona 08034, Spain}

\author{Mohammed Bouabdellaoui}
\affiliation{Aix Marseille Univ, Universit\'e de Toulon, CNRS, IM2NP 13397, Marseille, France}

\author{Marco Abbarchi}
\affiliation{Aix Marseille Univ, Universit\'e de Toulon, CNRS, IM2NP 13397, Marseille, France}

\author{Pietro de Anna}
\affiliation{Institute of Earth Sciences, University of Lausanne, Lausanne 1015, Switzerland}
\email{pietro.deanna@unil.ch}

\maketitle

\section*{Abstract}
\noindent
\textbf{Natural porous systems, such as soil, membranes, and biological tissues comprise disordered structures characterized by dead-end pores connected to a network of percolating channels. The release and dispersion of particles, solutes, and microorganisms from such features is key for a broad range of environmental and medical applications including soil remediation, drug delivery and filtration. Yet, the role of microscopic structure and flow for the dispersion of particles and solutes in such disordered systems has been only poorly understood, in part due to the stagnant and opaque nature of these microscopic systems. Here, we use a microfluidic model system that features a pore structure characterized by dead-ends to determine how particles are transported, retained and dispersed.  We observe strong tailing of arrival time distributions at the outlet of the medium characterized by power-law decay with an exponent of 2/3. Using numerical simulations and an analytical model, we link this behavior to particles initially located within dead-end pores, and explain the tailing exponent with a hopping and rolling mechanism along the streamlines inside vortices within dead-end pores. These dynamics are quantified by a stochastic model that predicts the full evolution of arrival times. Our results demonstrate how microscopic flow structures can impact macroscopic particle transport.} 

\section*{Introduction}
Most geological, biological and industrial systems are characterized by a porous structure where fluids can move through a network of small confined spaces (also known as pores) of size $\lambda$ that can vary within a broad range.  On one hand, the variability in pore-size induces velocity heterogeneity leading to anomalous transport~\cite{deAnnaPRF2017, datta2013, Dentz2018},  mixing~\cite{deAnnaEST2014,Heyman2020},  filtration~\cite{Nishiyama2012, miele2019}, and microbial dispersal \citep{Scheidweiler2020, deAnnaNaturePhys2021}. Anomalous transport has been also related to internal vortices observed next to pore-constrictions \citep{Cardenas2009}.  On the other hand, morphological diversity introduced by disordered pore structures \citep{Alhashmi2016, Xiong2016, Wu2019} also induces a rich flow organization with spatial and temporal complexities that play critical roles in natural and engineered phenomena, such as groundwater contamination and remediation~\cite{Kahler2019}, enhanced hydrocarbon recovery~\cite{Kar2015} and water filtration systems~\cite{Kosvintsev2002}. The complex morphology of a porous system often results in dead-end pores, the portion of the system that cannot host a net fluid transfer resulting in stagnant flow~\cite{Lever1985}. Because of this nature, it is common to assume that within these dead-end-pores, the transported quantities remain trapped for long times until molecular diffusion allows them to escape (a very slow and inefficient mechanism~\cite{Wirner2014, Battat2019}). However, a complete quantitative understanding of the role played by the local flow structures on the asymptotic transport associated to dead-end-pores remains elusive to date.

To unravel the link between microscopic motion within dead-end-pores and porous media transport, we consider a medium composed of a network of randomly distributed transmitting-pores (TP) and dead-end-pores (DEP). To build such a complex structure, we exploit the novel method of spinodal solid-state dewetting~\cite{salvalaglio2020}: it is based on annealing of mono-crystalline semiconductor thin films that results in surface morphologies exhibiting random topology and a hyperuniform character, see Fig. \ref{fig1}$a, \, b$. Such a dewetted geometry represents the impermeable backbone of a relatively homogeneous porous medium composed of TP (about 94~\% of the total volume) and DEP (about 6~\% of the total volume) with structural disorder that is statistically isotropic,  similar to crystals or jammed disordered spheres packing~\cite{Zachary2011}. Despite its overall homogeneity in terms of TP-size $\lambda$ (\textit{i.e.} the distribution of $\lambda$-values is very narrow; see Fig.~\ref{fig1}$c$), we show that this porous structure has distributed DEP-depth (see Fig. \ref{fig1}$d, \, e$ and $f$) and displays complex flow and transport properties that directly link to the underlying porous structure.

\section*{Results}

\subsection*{Medium characterization}
\begin{figure}[htb!]
	\centering
	\includegraphics[width=1\linewidth]{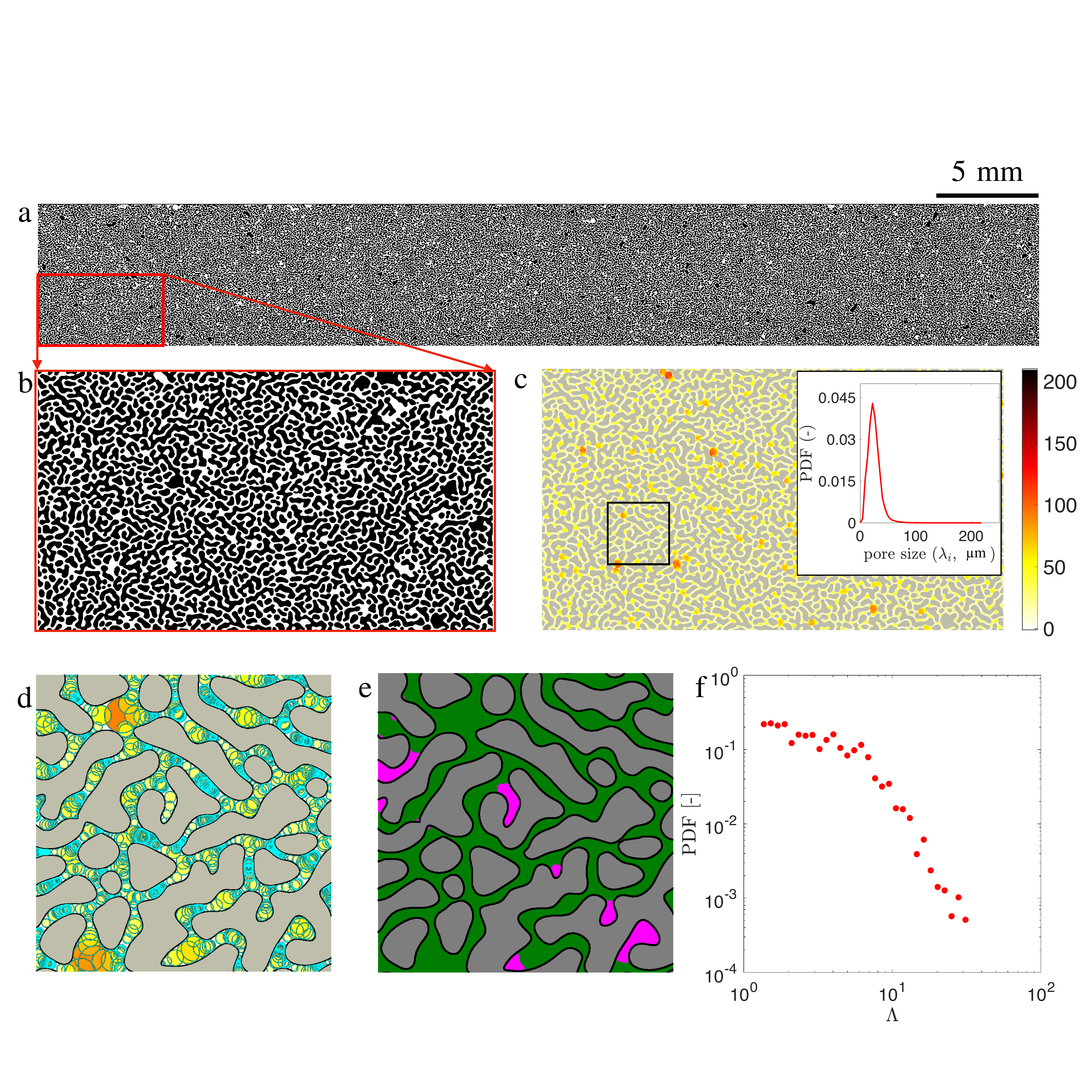}
	\caption{\textbf{Characterization of the model porous structure reveals dual feature of the complex medium. a,} Binary image of the disordered hyperuniform porous structure, and \textbf{b,} a close-up view. The disordered hyperuniform structure exhibits complex porous network (white) interspersed among disordered grains (black) within the system. \textbf{c,} A map of pore size ($\lambda$, [mm]) and its probability density function, PDF (in inset), which is narrow about the average value $\lambda_m$ = 0.027 mm. \textbf{d,} A further enlarged view highlighting the inscribed circles (cyan) along the porous network that estimate the local pore-size (see Methods $a$). The colorbar in \textbf{c} applies to \textbf{d}. \textbf{e,}  The dual feature of the medium characterized by the transmitting-pores (TP: green) and the dead-end pores (DEP: magenta). The two features are segregated based on number of grains ($\zeta$) surrounding a pore-space, such that DEP has $\zeta=1$, and TP has $\zeta>1$. \textbf{f,} The PDF of effective aspect ratio of a DEP, defined as the ratios between the area of a DEP ($\mathcal{A}$) and the mean pore-space area ($\lambda_m^2$).}
	\label{fig1}
\end{figure}

We use soft lithography \cite{rusconi_microfluidics_2014} to print the hyperuniform porous structure (HPS) (Figs.~\ref{fig1} $a$, $b$) into a silicon wafer that acts as a mold for a PDMS microfluidic chip (see SI). The irregular grain features of the HPS represent the solid matrix of the porous microchannel, characterized by a porosity $\phi = 0.39$. Between the inlet and outlet zones, the microfluidic channel has length $L=50$~mm, width $w=7$~mm and thickness $a=0.083$~mm, schematically shown in Fig. \ref{fig2}$b$. Its thickness $a$ is designed to be of the same order of magnitude as the average pore size, $\lambda_m$= 0.027 mm, yielding velocity profiles similar to square-channel flow in the TP, as verified by PIV (see SI Fig. 1). The local pore size $\lambda$ is computed using the method of maximum inscribed circle (MIC, see Methods $a$) across the whole domain. The medium is quite homogeneous, such that the statistical distribution of $\lambda$ is narrow and has a strong peak close to the mean pore-size $\lambda_m = 0.027$~mm (inset in Fig.~\ref{fig1}~$c$), representative of many common geological structures~\cite{bear_dynamics_1988}. We characterize the porous structure in terms of the segregation index $\zeta$ (see Methods $a$ and Figs.~\ref{fig1}$c, d$) to distinguish between TP and DEP, respectively shown as green and magenta regions in Fig.~\ref{fig1}$e$. It captures the dual feature of the matrix (TP: $\zeta > 1$ and DEP: $\zeta$ = 1), such that $\zeta$ refers to the number of individual grains surrounding each pore. The DEPs are the elongated pores caving inside a singular grain in contrast to the TPs that are surrounded by multiple grains. We quantify the aspect ratio $\Lambda$ of each DEP as the ratio between their area $\mathcal{A}$ and $\lambda_m^2$: it results that $\Lambda = \mathcal{A}/\lambda_m^2$ spans within the range $1-30$ (Fig.~\ref{fig1}$f$). 

\subsection*{Experiments}
\begin{figure}[htb!]
	\centering
	\includegraphics[width=1\linewidth]{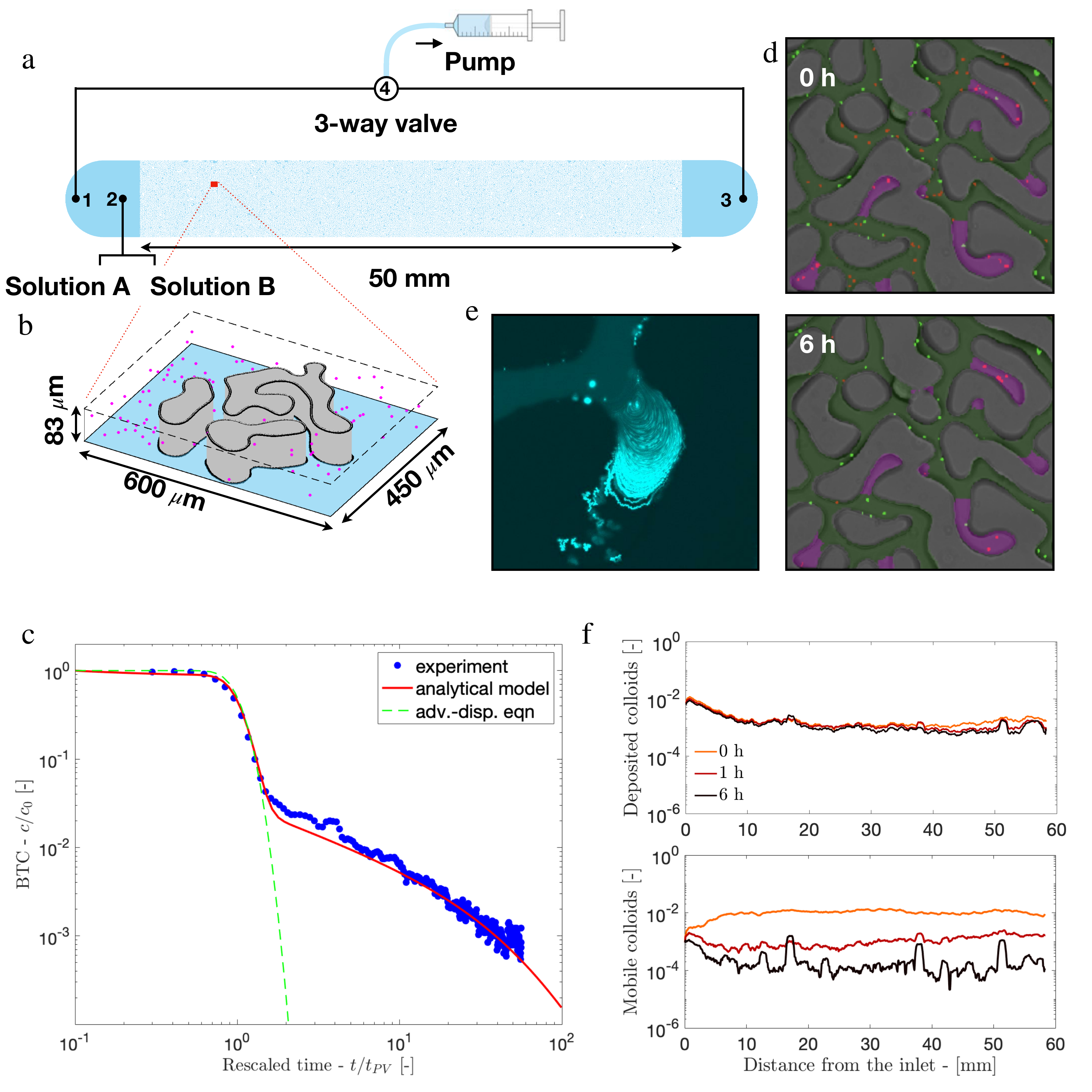}
	\caption{\textbf{Dual geometric feature of the medium leads to two distinct regimes in the breakthrough curve of colloidal particles.} \textbf{a,} Schematic of the experimental setup and \textbf{b,} of a single dead-end-pore. \textbf{c,} Experimental breakthrough curve (BTC, blue dots) computed as the $c(t)/c_0$, where $c_0$ is the injected colloidal numerical density, and $c(t)$ is the measured density eluted at time $t$. The dashed line represents the analytical solution of advectiion-dispersion equation (equation 26 in SI). The solid red line represents the prediction of the analytical (CTRW) model (equation \ref{eq:BTCall}). \textbf{d,}  Two snapshots of mobile (red) and immobile (green) colloids initially and 1 hour after injection;  respectively; TPs are represented as green areas while DEPs as magenta. \textbf{e,} Qualitative identification of vortex structure inside a DEP captured by time-stacking images taken  at $Pe \sim 10^5$. \textbf{f,} Profile of deposited (above) and mobile particles (below) along the channel length acquired at three different times (0, 1 and 6 hrs) after the start of the experiment.}
	\label{fig2}
\end{figure}

We investigate the impact of this dual feature of the medium (namely, TP and DEP) on the macroscopic transport of a neutrally buoyant suspension of colloidal particles. To this end, we first produce a {near-}homogeneous distribution of suspended colloids across the porous medium by continuously injecting a density matched suspension ($A$, polystyrene micro-spheres of diameter $0.5 \, \mu$m and density $\rho = 1.05$~g/mL Thermofischer Fluoromax B500 in 1:1 milliQ-D$_2$O mixture) within the microchannel at a flow rate of $Q = 0.2~\mu$L/min for 24 hours. Using the same strategy to inject a sharp front as in~\cite{deAnnaNaturePhys2021}, a displacing solution with the same density ($B$, 1:1 milliQ-D$_2$O mixture), is then withdrawn at a constant flow rate $Q = 0.5 \, \mu$L/min through the breakthrough circuit (see \ref{fig2}~$a$ B-2-3-4) eluting one pore volume (i.e. the volume of the entire porous channel) every $T_{PV} = 21.8$~min, for about 60 $T_{PV}$. The relative importance of advection to diffusion process is quantified by the corresponding P\'eclet number, $Pe = \lambda_m u_m/D$ based on the mean fluid velocity ($u_m = 0.038$ mm/s: the flow regime is laminar since $Re = u_m \, \lambda_m / \nu \sim 0.001$) and the suspension diffusion coefficient, $D = 1.4 \times 10^{-7}$~mm$^2$/s, is $Pe = 7329.$ We independently measure the diffusivity of 0.5~$\mu$m polystyrene colloids in the studied medium and find $D$ to be  smaller than the theoretical prediction for suspension in the bulk from the Stokes-Einstein equation due to the confined structure of the system (see SI).
Time-lapse composite imaging coupled with fluorescence microscopy (see Methods $b$) allows us to count the number of effluent colloids near the outlet and measure the breakthrough curve (BTC) over three orders of magnitude. The latter, displayed in Fig.~\ref{fig2}$c$, shows two distinct transport regimes. The first regime, limited to the elution of about two pore-volumes, is well captured by the classical advection-dispersion framework~\cite{Dentz2011} (see Methods $d$). The dispersion coefficient has been fitted and results to be $D^\ast = 0.03$~mm$^2$/s (green dashed line in ~\ref{fig2}$c$) using a least-squares method (see Methods $d$). At later times ($t/T_{PV} > 2$), a second regime emerges, where the BTC shows a power-law-like heavy tail, reminiscent of the so-called anomalous transport behavior (e.g.~\cite{Dentz2011}). 

We also monitor the spatial distribution of suspended colloids within the porous system (see Fig.~\ref{fig2}$f$) by periodically collecting two composite images of the entire microfluidic channel (composing $38 \times 9$ individual pictures) 7 times (0, 1, 4, 10, 16, 22 and 44 hour after injection), temporally separated by $\Delta t = 6$~min. By comparing each couple of consecutive images, we distinguish the mobile particles from those retained by the host solid structure (see Methods $e$). Fig.~\ref{fig2}~$f$ shows that the spatial profiles of the retained particles (top) at three different times ($t=0, \, 4$~h, $8$~h) remain constant (thus, no deposition takes place during the flow experiment), while those of mobile particles (bottom) decay with time homogeneously across the porous channel. 

Because the fluid velocities within the DEPs are significantly smaller than the ones within the TPs (as also verified by numerical simulations Fig.~\ref{fig3}$c-d$), it is common to assume that the DEPs only delay the macroscopic transport by trapping the solute/suspensions that get released by diffusion. As a consequence, porous systems comprising of DEPs are generally modeled as  dual media with separate advection-dominated and diffusion-dominated regions (e.g. \cite{Gouze2008}). These models predict that particles move fast through the advection-dominated region, or they spend long time trapped in a diffusion-dominated region, and their transition across the two regions is characterized by a given transfer rate. Contrary to this simplified view about such systems, we show the existence of complex vortex flow structures within the DEPs that, coupled with molecular diffusion, control the macroscopic BTC. To qualitatively show these flow structures, Fig.~\ref{fig2}$e$ displays time-stacking of transported colloidal particles for $Pe \sim \mathrm{10^5}$ at the entrance of a DEP, where the velocity is smaller than in the adjacent TP, a laminar vortex is identified by closed trajectories. Far inside the DEP, the individual tracks of random walkers are visible as short and tortuous segments suggesting the dominance of molecular diffusion. 

\subsection*{Numerical simulations}
\begin{figure}[htb!]
	\centering
	\includegraphics[width=1\linewidth]{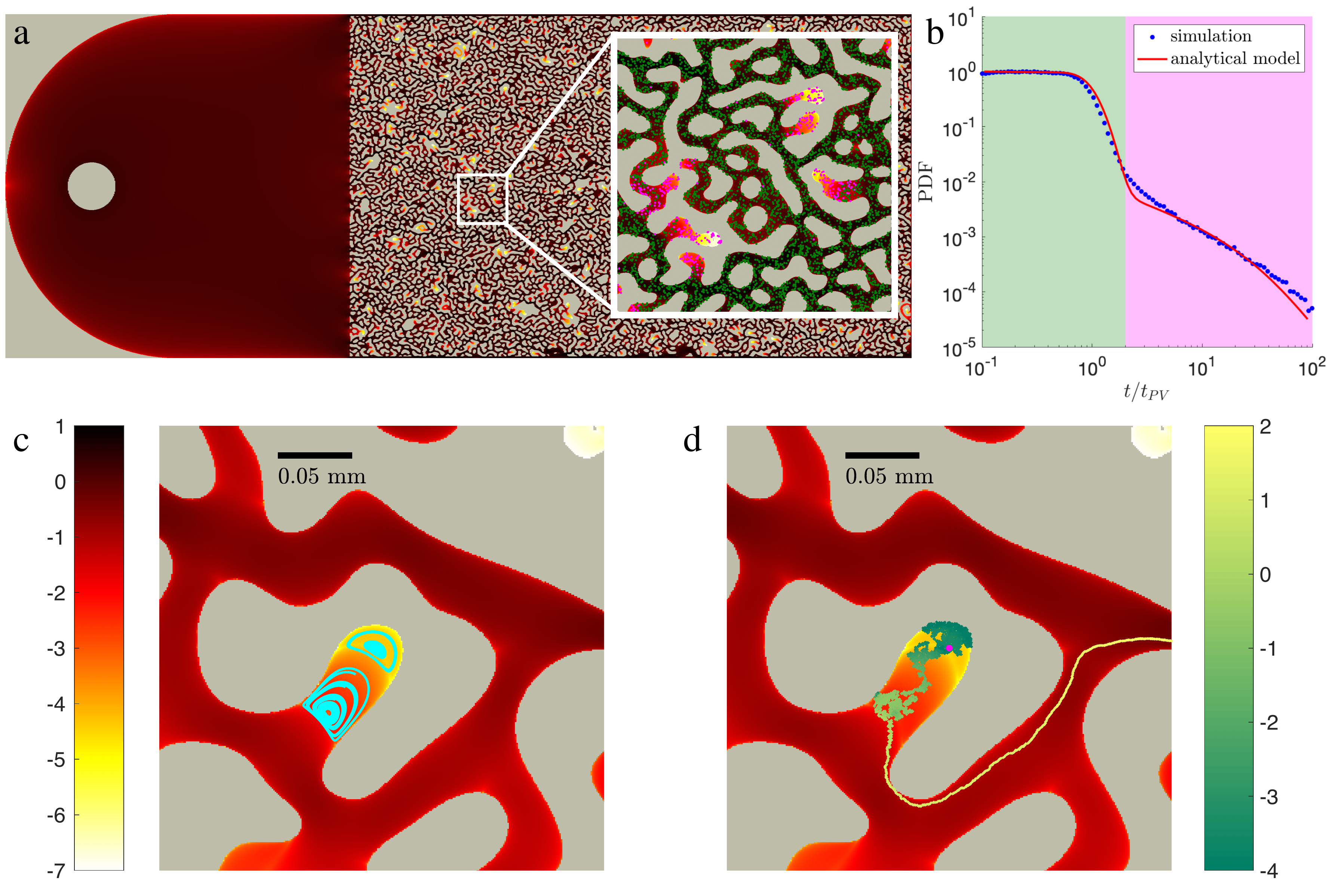}
	\caption{\textbf{Computation of velocity field and particle tracking simulation through hyperuniform structure:} \textbf{a,} The modulus of the Stokes flow solutions (mm/s in log-scale) in a subsection ($\mathrm{1/5^{th}}$ in length) of the porous medium used in the experiment is superposed to particles that initially occupy the TP (green) and DEP (magenta) (enlarged view in the inset); \textbf{b,} Probability density function (PDF) of particle escape time (equivalent to the BTC) versus normalized time ($t/t_{PV}$) obtained from particle tracking in the simulated velocity field (blue dots) and the analytical  (CTRW) model, equation \ref{eq:BTCall} (red line) with $\alpha = 0.067$. The magenta and green shades distinguish the regions of the BTC contributed by the particles shown in corresponding colors in the inset of \textbf{a}. The long tail of the PDF is contributed by particles originating in the DEPs; \textbf{c,} close view of the vortex structures inside a DEP; \textbf{d,} an individual trajectory of a particle originated at the magenta dot and leaving the DEP color-coded with local P\'eclet number $Pe^* = \lambda_mv_p/D_m$ in log-scale.}
	\label{fig3}
\end{figure}

The structure of a cavity flow, similar to the one observed within a DEP, is characterized by a cascade of vortices with exponentially decaying velocity magnitude that spans a very broad range of scales. Therefore, a detailed understanding of such velocity field requires a multi-scale description. Unfortunately, it is very challenging to resolve the multi-scale structure of such flow-fields using the state-of-the-art experimental techniques, such as PIV (see SI: Fig. 1). Herein, we use COMSOL Multiphysics to numerically solve the two-dimensional steady state incomprehensible Stokes flow equations in a subsection of the microfluidics geometry (see Methods $f$ and SI). The domain of this numerical computation is approximately one-fifth in length (11~mm) and the same in width (7~mm), as that in the experiment. The computed velocity magnitude, shown in Fig.~\ref{fig3}~$a$ in logarithmic scale (increasing velocity from light to dark). Further, we simulate the Lagrangian trajectories of $N = 10^5$ particles {initially distributed homogeneous across the medium and} transported by {a combination of the} computed velocity field and molecular diffusion with $D = 7.9 \times10^{-7}$~mm$^2$/s (see Methods $f$), using the Langevin equation:
%
	\begin{align}
		\vec{x} (t+\Delta t) = \vec{x}(t) + \vec{v}[\vec{x}(t)] \Delta t +\sqrt{2D\Delta t} \boldsymbol \xi(t),
	\end{align}
%
where $\vec{x}(t)$ is the particle position at time $t$, $\vec{v}(\vec{x})$ is the local flow velocity (obtained by interpolating the computed velocity values to the particle location \citep{deAnnaPRF2017}), and $\xi(t)$ is a Gaussian random number with zero mean and unit variance. We compute the BTC of the simulated particles (see Fig. \ref{fig3}$b$) that shows the same two distinct regimes observed in our experiment. The late-time tail of the BTC is depends on the global P\'eclet number ($Pe$), such that an increasing $Pe$ results in a longer tail (see SI). We investigate further into the two regimes (shown in green and magenta shades in Fig. \ref{fig3}$b$) by finding the initial positions of the particles with the same colors in the inset of  Fig.~\ref{fig3}$a$. It is clear that the late time scaling in the BTC is due to the particles initially trapped within the DEPs, whereas the particles not in the DEPs are washed-out from the medium before the transition time, about $t/T_{PV} = 2$. A close-up view of the velocity field magnitude inside a DEP (see Fig.~\ref{fig3}$c$) superposed with a few characteristic streamlines (in cyan) reveal the underlying flow structure organized in a cascade of laminar vortices. 

The spatial structure of fluid velocity along the depth of a DEP is similar to that of one of the classical cavity flows~\cite{shankar1993}. In this scenario, a particle can escape a vortex of closed streamlines only by hopping across them via molecular diffusion. This mechanism is demonstrated in Fig. \ref{fig3}$d$ by plotting a single trajectory initiated deep inside a DEP and color-coding it with the particle P\'eclet number (increasing in values from dark to light) defined as $Pe^{*} = \lambda_m v_p / D$, where $v_p$ is the local velocity of the tracked particle. Initially being in the low velocity zone ($v_p\sim10^{-7}$ mm/s), the particle is transported randomly across streamlines by molecular diffusion ($Pe^{*}\sim10^{-4}$), and as it approaches the outer vortex it hops across streamlines but also rolls along the vortex ($Pe^{*}\sim 1$), before being advected away by the TP flow ($Pe^{*}\sim10^{2}$). 

\subsection*{Theoretical analysis}
\begin{figure}[htb!]
	\centering
	\includegraphics[width=1\linewidth]{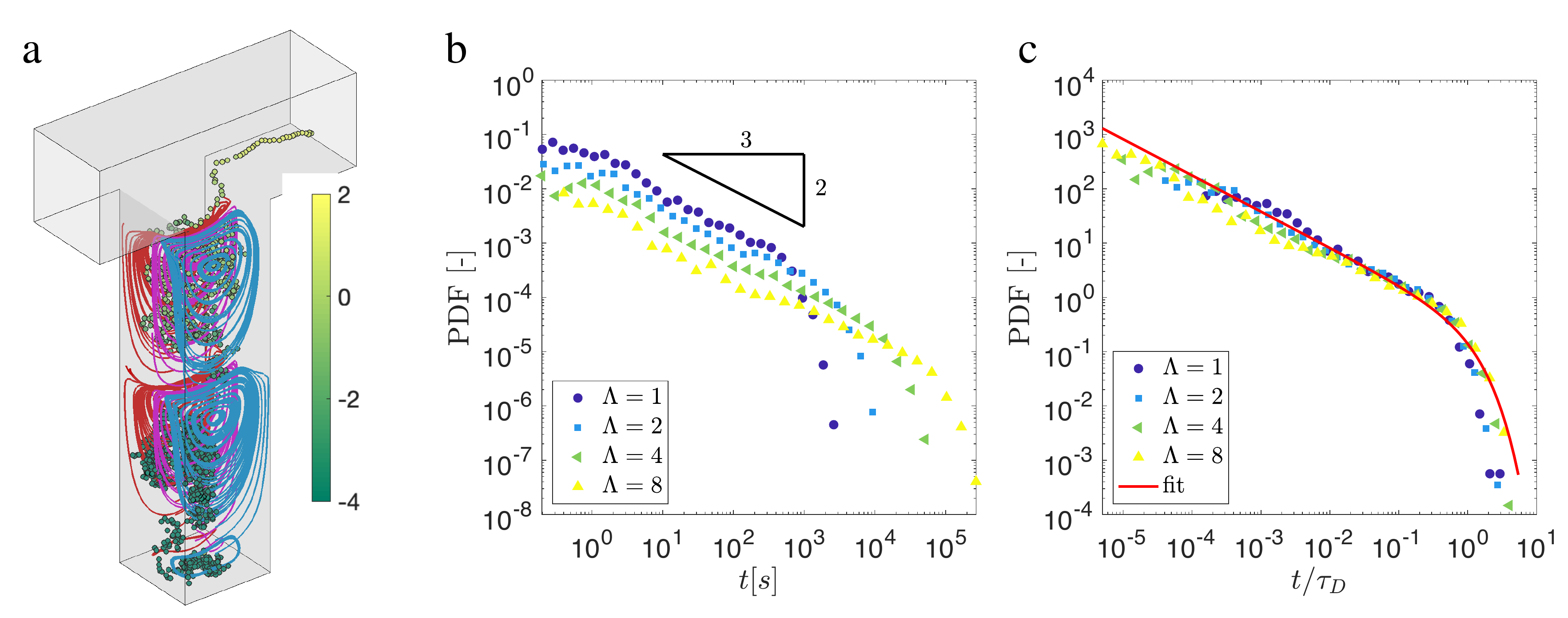}
	\caption{\textbf{Conceptual model for DEP flow capturing the local particle escape time}: \textbf{a,} A single numerically simulated trajectory (color-coded with local P\'eclet number $Pe^* = \lambda_mv_p/D_m$ in log-scale) originating at the bottom of a  3D rectangular cavity representing a DEP (aspect ratio, $\Lambda$ = 4) connected to a square channel representing a TP.  A series of streamlines highlights the vortex flow structure inside the cavity. \textbf{b,} Particles escape time probability density function (PDF, equivalent to their BTC) of a single cavity for $\Lambda = 1, 2, 4, 8$, and \textbf{c,} the same as \textbf{b} with time rescaled by diffusion time-scale $\tau_D = (\lambda_m\Lambda)^2/D$.}
	\label{fig4}
\end{figure}

To understand how this mechanism of rolling and hopping across streamlines in a vortex controls the late time behavior in the BTC, we first model the transport in a single DEP generalizing its geometry to a cavity connected to a free channel. The widths of cavity and free channel are set to be equal to $\lambda_m$, and the cavity-depth is varied based on four depth-to-width aspect ratios $\Lambda = 1, \, 2, \, 4, \, 8$. We perform this simulation for both two- and three-dimensional geometries, adding a uniform thickness of $\lambda_m$ in the latter. Using the same numerical scheme adopted to solve the flow in Fig.~\ref{fig3}, we compute $\mathrm{10^4}$ particle-trajectories initiated homogeneously inside each cavity volume. Fig.~\ref{fig4}$a$ shows the result for a 3-dimensional representative case with $\Lambda = 4$: the flow structure is highlighted by multiple streamlines (initiated at three $y$-plane locations shown in different colors) associated with the vortices of decaying intensity along the depth~\cite{shankar1993}. The trajectory of a particle initiated at the cavity bottom is shown with the same color-scheme as in Fig. \ref{fig3}$d$. The trajectories for both 3D and 2D cavities (see also SI: Fig. 5) show behavior similar to the ones in the DEP of the porous medium: the particle initially diffuses isotropically ($Pe^{*}\ll1$) until it reaches the upper part of the cavity, where its motion is a result of the competition between advection along the vortex streamlines and diffusion that promotes streamline exchange ($Pe^{*} \sim 1$). Finally, the particle exits the cavity and follows the channel (TP) flow with $Pe^{*} \gg 1$. 

Next, we compute the BTC for the four tracking simulations, $\Lambda = 1, \, 2, \, 4, \, 8$, as the PDF of particle arrival times {to the channel outlet}. Fig.~\ref{fig4}$b$ shows that, for all cases the distribution decays as the power law $t^{-2/3}$ with an exponential cut-off: a Gamma distribution. {This power-law exponent is different from the scaling expected for diffusion alone, $t^{-1/2}$}~\cite{Grebenkov2019}. The cut-off time for each $\Lambda$ is given by the characteristic diffusion time over the DEP depth, such that  $\tau_D = (\lambda_m \Lambda)^2/D$. Hence, when time is rescaled by $\tau_D$, the BTCs for all aspect ratios $\Lambda$  collapse into the master-curve
%
	\begin{equation}\label{eq:BTC}
		g(t) = \frac{(t/\tau_D)^{-2/3} \exp(-t/\tau_D)}{\Gamma(1/3)}.
	\end{equation}
%
The characteristic scaling of $t^{-2/3}$ is controlled by the trapping of particles in vortex streamlines. The combined action of advection-controlled transport in the shear flow along the solid boundaries within a vortex, and diffusion at the open boundaries of the vortex itself leads to a scaling of residence times as $\psi(t) \sim t^{-1-\gamma}$ with $\gamma = 2/3$, as discussed in a different context by~\cite{Bouchaud1990}. This residence time distribution corresponds to a trapping time distribution scaling as $t^{-\gamma}$. A detailed derivation of this scaling exponent is provided in the SI.  

Based on the above scaling, we formulate an analytical model (see SI) by considering the porous system as combination of several DEPs with an aspect ratio $\Lambda$ that is distributed as in Fig. \ref{fig1}$e$, resulting in a distributed time scale $\tau_D$. The global BTC can be constructed as the weighted average between the BTC of particles originated within TPs (solution of advection-disperion equation) and the BTC of particles orginated within DEPs. The latter is obtained from the above Gamma distribution (equation \ref{eq:BTC}) weighted by the probability density function $f_D(t)$ of the corresponding characteristic diffusion time $\tau_D$,
	\begin{align}\label{eq:BTCall}
		F(t) = (1 - \alpha) \frac{1}{L} \int\limits_0^L dx f_0(t,x) + \alpha \int\limits_0^\infty d \tau g(t/\tau) \tau^{-1} f_D(\tau).
	\end{align}
%
Here, $\alpha$ is the fraction of observed colloids initially located in DEP and $1-\alpha$ the fraction of those located in TPs. The function $f_D(t)$ can be expressed in terms of the measured PDF $f_\Lambda(a)$ of aspect ratios $\Lambda$ as $f_D(t) = f_\Lambda(\sqrt{t D/\lambda^2})\sqrt{D/\lambda^2 t}/2$, and $f_0$ is the solution of the advection-dispersion equation representing particle originating in TPs (see Methods and SI). The parameter $\alpha$ for the experiment is measured as 0.22 (see SI) and that for the simulation is set as 0.067 for the non-homogeneous and homogeneous colloid distribution, respectively (we performed other simulations varying $\alpha$, see SI). The predictions of our model in eq.~\eqref{eq:BTCall}, which are completely based on the medium geometry and the flow characteristics, are shown alongside the experimental BTC (Fig.~\ref{fig2}$c$) and the BTC from numerical simulation (Fig.~\ref{fig3}$b$). The predictions of the model for our simulation with varying P\'ectlet number are shown in SI. This analytical model fully captures the early time and anomalous late time scaling, as well as the transition between the two regimes, of the measured and simulated BTC, with the macroscopic dispersion coefficient $D^\ast$ being the only fitting parameter. 

\section*{Conclusions}

These findings shed new light on the fundamental mechanism governing particle dispersion in disordered porous systems characterized by dead-end pores of fluid stagnation. Classical descriptions overlook dead-end flow structures assuming that fluid stagnation does not play a significant role on macroscopic transport~\cite{Benichou2008, Wirner2014}. Such diffusion based models cannot quantitatively predict the observed $2/3$~power-law decay of the macroscopic BTC. The observed microscopic flow within the dead-end pores is vortical, and it governs the late arrival time distribution. Here, in the case of colloidal transport, we have shown a quantitative link connecting the characteristic decay of macroscopic BTC to the delay associated with a hopping and rolling mechanism along the streamlines inside the dead-end pores. The particles diffusivity and the size distribution of dead-end pores control the BTC long tail cutoff, but they do not change the observed power-law scaling. Since the model described here is not dependent on the system dimensionality, it can be readily applicable to disordered porous systems encountered in natural soil and other environmental systems. Given the ubiquity of dead-end structures with local flow recirculation in such systems, we anticipate that the mechanism and model proposed here would be useful to understand the role of transport through brain extracellular spaces~\cite{Nicholson2017} and applications related to drug delivery, or phenomena taking place at very different scales such as Karst conduits~\cite{Hauns2001}. 

\section*{Methods}
\subsection*{Medium characterization}
To obtain maps of the pore-size ($\lambda$) distribution and the segregation index ($\zeta$) of the model porous matrix, we analyzed the image representing the medium geometry (Fig.~\ref{fig1}~$a$). This image is a binary two-dimensional array of pixels that distinguishes the pore space (pixel value = 1, white) from the grains (pixel value = 0, black). First, we labeled each grain and identified its boundary. The binary image was then skeletonized to obtain the 1-pixel width representation of the pore space (see SI Figs. 2). For a specific skeleton location, we employed a maximum inscribed circle (MIC) algorithm (see SI) that fits the largest circle on its neighboring grain boundaries based on the Euclidean distance map. By iteratively scanning all skeleton locations with this algorithm, we generated the series of the locally largest circles touching the grain boundaries at three points and spanning across the entire pore space domain. The pore size ($\lambda$) at a specific location was assigned as the average diameter of the overlapping circles around that location. The segregation index ($\zeta$) was assigned at each location as the number of individual, different, grains touched by the inscribed circle at that location. A pixel with $\zeta$=1 belongs to a dead-end pore and that with $\zeta > 1$ belongs to a transmitting pore (TP). For locations outside the fitted circles the segregation index was obtained via linear 2D interpolation of the neighboring $\zeta$ values.

\subsection*{Time-lapse video-microscopy} 
Time-lapse imaging was performed with an automated inverted microscope (Eclipse Ti2, Nikon) equipped with a CMOS camera (Hamamatsu ORCA flash 4.0, 16-bit). All individual pictures ($2048 \times 2048$ pixels) were recorded at $10$X magnification (corresponding to $0.65\, \mu$m/pixel) focusing the microscope optics on the middle horizontal plane of the microfluidic channel. Colloids were imaged by fluorescence microscopy (using a Nikon DAPI fitler cube) with an exposure time of $50$ ms. Pictures of the flowing colloids were recorded every $5$ minutes close to the outlet by a composite image of $9$ individual pictures along the transverse flow direction to cover the entire cross section of the channel outlet.

\subsection*{Measuring BTC} 
In order to compute the number $n$ of effluent colloids at the time $t_k$, we first accounted for the immobile particles detected in the field of view by removing the image at time $t_{k+1}$ from the image at time $t_k$, we, then, applied a bandpass filter with a characteristic noise length of 1 pixels to every recorded image in order to smooth the electronic noise associated with camera acquisition. We identified the colloids as individual peaks with a minimum brightness of 5\% the pixel depth (16 bit) and minimum diameter of 3 pixels. The BTCs of effluent colloids were obtained as $c = n(t_k) / n_0$, normalizing the measured number of colloids eluted at time $t_k$ by the count $n_0$ at the beginning of the experiment ($t_0$). The physical time $t$, defined as time elapsed since the injection begins, was rescaled by the residence time of one pore volume, defined as $T_{PV} = L \, w \, a \, {\phi} / Q = 21.8$ min, where $L = 51$ mm is the longitudinal size of the porous system, $w = 7$ mm is channel width, $a = 0.083$ mm is its thickness, $\phi = 0.39$ is the medium porosity and $Q = 0.5$ mm$^3$/min is the imposed flow rate, corresponding to an average (Darcy) velocity of $q = 0.038$~mm/s.

\subsection*{Classical transport model through homogeneous porous media}
The classical model to describe transport through a relatively homogeneous porous medium is the so-called advection-dispersion framework that expressed mass conservation as~\cite{bear_dynamics_1988,Dentz2011}:
%
	\begin{equation}\label{eq:ADE}
	\frac{\partial c}{\partial t} = - q \, \frac{\partial c}{\partial x} + D^\ast \frac{\partial^2 c}{\partial x^2}
	\end{equation}
%
where $D^\ast$ is the macroscopic dispersion coefficient, in mm$^2$/s. Our experimentally measured BTC are well matched by the analytical solution $c(x,t)$ of the above equation (ADE) evaluated at the medium outlet $x = L$ mm~\cite{deAnnaNaturePhys2021} (Fig~\ref{fig2}~2$a$ green dashed line) for times shorter than about 2 $T_{PV}$. For larger times the TP are basically empty and only DEP mass release contribute to the BTC that deviates from exponentially decaying behavior to scale as a heavy-tailed power law. 

\subsection*{Spatial organization of the suspended and deposited colloids}
Profiles of deposited and mobile colloids were computed from the composite images of the entire microfluidic channel at 0, 1, 4, 10, 16, 22 and 44 hours. Clusters of connected pixels larger than $3 \times 3$ pixels were considered as colloids after removing background via an adaptive-thresholding algorithm which chooses a threshold value based on the local mean intensity over an area of $1001 \times 1001$ pixels (using the Matlab embedded function \textit{adaptthresh}). Colloids detected in consecutive images with overlapping positions have been considered as deposited (Fig.~\ref{fig2}~$d,e$ green dots), while the non-overlapping colloids were categorized as mobile (Fig.~\ref{fig2}~$d,e$ red dots). The local surface occupied by these two classes of colloids has been integrated along transversal slices of $10 \, \mu$m and normalized by the accounted porous area to measure the mobile and deposited profiles, shown in Fig.~\ref{fig2}~$f$. During the experiment, deposited colloids profile are not changing, meaning that the deposition that took place during the saturation process does not affect the transport experiment.

\subsection*{Numerical simulation to predict colloid transport}
The Stokes flow solution used to model the colloids transport is computed over a domain discretized using a physics based unstructured mesh with approximately 5$\times10^{6}$ elements and the element size adapting to the geometry with a minimum of 0.2 $\mu$m (see SI: Fig. 4). We imposed no-slip boundary conditions at all grain and domain boundaries, a zero reference pressure at the outlet and an uniform flow rate $Q = 0.5~\mu$L/min at the inlet. The computational resolution is high enough to ensure a divergence free velocity field. 

\newpage
\newpage
\newpage
\newpage

\noindent
\textbf{Additional information}\\
A detailed description of the microfluidics fabrication and derivation of the characteristic scaling laws can be found in the Supplementary Information.\\

\noindent
\textbf{Acknowledgments}\\
This work received the support of FET-Open project NARCISO (ID: 828890) and of Swiss National Science Foundation (grant ID 200021 172587). The authors also acknowledge the assistance from Faderico Pasotti in performing the PIV experiment, Silvia Guadagnini in measuring the diffusivity of colloids and Monica Bollani for very useful discussions.

\noindent
\textbf{A.D.B., D.S. and P.d.A. designed the research, M.A. and M.B. prepared the de-wetted HPS samples, D.S. performed transport experiments, A.D.B. performed numerical simulations, A.D.B., D.S. and P.d.A. analyzed the data, M.D., A.D.B., D.S., and P.d.A. derived the theoretical model, and all authors wrote the manuscript. A.D.B. and D.S. share the first authorship.} \\

\noindent
\textbf{Competing financial interests}\\
The authors declare no competing financial interests.

\subsection*{References}
\bibliographystyle{unsrt}
\bibliography{library}

%
%

\end{document}


\title{SUPPLEMENTARY MATERIAL FOR\\
Structure induced vortices control anomalous dispersion in porous media}

\author{Ankur Deep Bordoloi}
\affiliation{Institute of Earth Sciences, University of Lausanne, Lausanne 1015, Switzerland}

\author{David Scheidweiler}
\affiliation{Institute of Earth Sciences, University of Lausanne, Lausanne 1015, Switzerland}

\author{Marco Dentz}
\affiliation{Spanish National Research Council (IDAEA-CSIC), Barcelona 08034, Spain}

\author{Mohammed Bouabdellaoui }
\affiliation{Aix Marseille Univ, Universit\'e de Toulon, CNRS, IM2NP 13397, Marseille, France}

\author{Marco Abbarchi}
\affiliation{Aix Marseille Univ, Universit\'e de Toulon, CNRS, IM2NP 13397, Marseille, France}

\author{Pietro de Anna}
\affiliation{Institute of Earth Sciences, University of Lausanne, Lausanne 1015, Switzerland}
\email{pietro.deanna@unil.ch}

\maketitle

\maketitle

\section*{Microfluidics fabrication and flow system}
Molds were fabricated by depositing a layer of SU-8 2150 (MicroChem Corp., Newton, MA) with controlled thickness of 0.083 mm on silicon wafers via spin-coating and the desired geometry was engraved via photolithography. Polydimethylsiloxane (PDMS; Sylgard 184 Silicone Elastomer Kit, Dow Corning, Midland, MI) was prepared with an addition of 10\% by weight of curing agent and casted on the molds. PDMS microchannels were plasma-sealed onto $25 \times 75$~mm glass slides~\citep{hol_zooming_2014}.

\section*{Diffusivity measurement in confined medium}
\begin{figure}
	\includegraphics[angle = 0,width=1\linewidth]{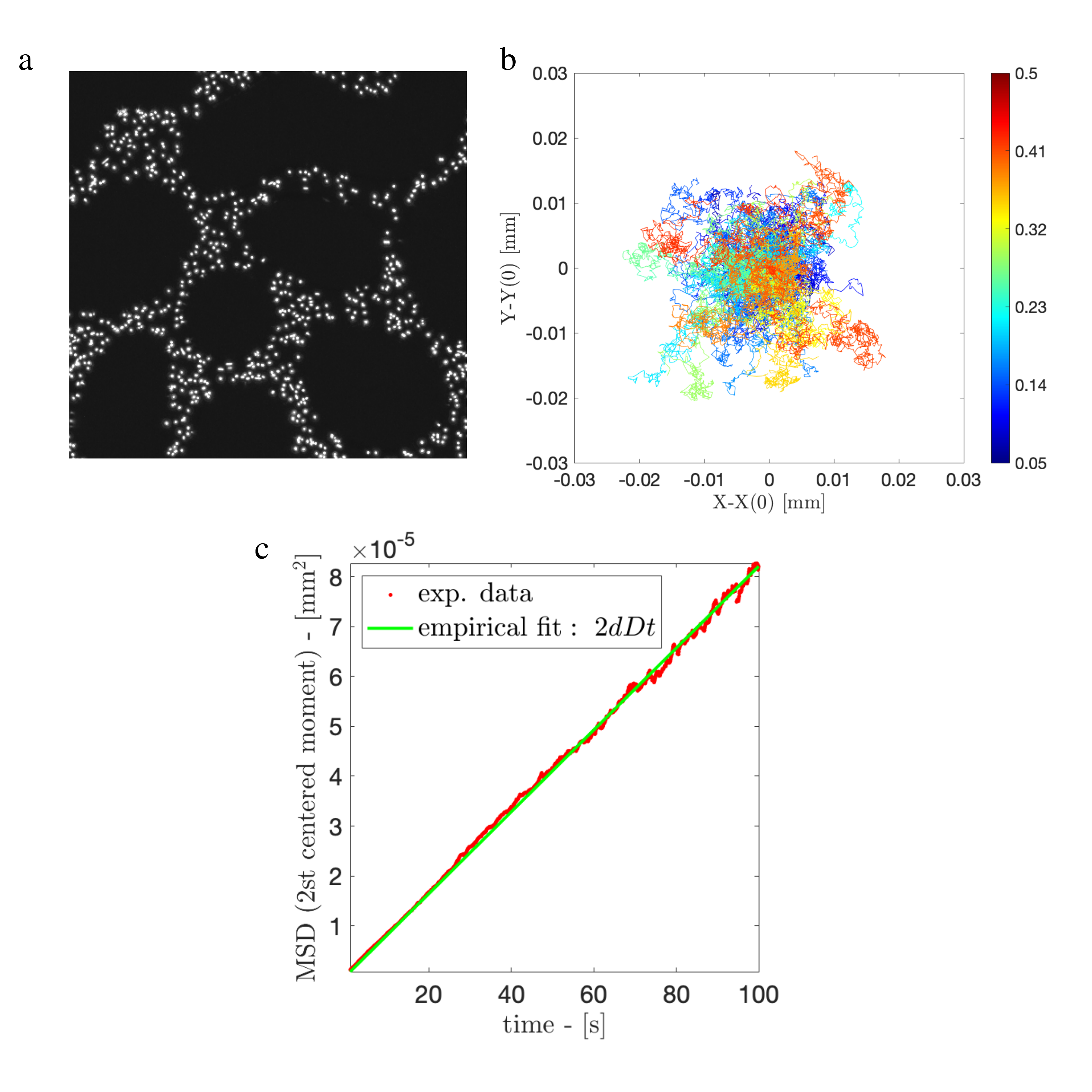}
	\caption{\textbf{Diffusivity of suspended colloids in confined hyperuniform porous medium:} \textbf{a,} Fluorescence microscopy image of 0.5 $\mu$m DAPI particles; \textbf{b} sample trajectories subtracted by their initial coordinates at $t=0$ and color-coded with the cumulative displacement (mm); \textbf{c} Temporal evolution of the mean-square displacement (MSD) averaged over 7296 trajectories. The fitted line ($2dDt$, where $d=3$ is the dimensionality of the system) measures the diffusion coefficient as, $D = 1.4\times10^{-7} \mathrm{mm^2/s}$.  \label{sfig6}}
\end{figure}

We independently measure the diffusion coeffcient ($D$) for 0.5 $\mu$m DAPI colloids in the same confined geometry used in the experiment tracking suspended particles under no flow condition. We capture a time sequence of fluorescence images of suspended particles at $\mathrm{1/10^{th}}$ of a second for 100 seconds. A small subsection of visualized particles is shown in Fig. \ref{sfig6}$a$. The sample trajectories shown in Figure \ref{sfig6}$b$ suggest that the particles have nearly negligible mean displacement. The diffusion coefficient is computed based on 7296 independent trajectories from the slope of the mean-square displacement $\mathrm{MSD} = \left<(r-\left<r\right>)^2\right>$, which is expected to scale as $2dDt$, where $r=\sqrt{(X-X(0)^2+(Y-Y(0)^2}$  and $d=3$ is the dimensionality of the system (see Figure \ref{sfig6}$c$).

The diffusion coefficient of a colloidal particle in the bulk of a fluid can be estimated using the Stokes-Einstein equation:
\begin{align}
D = k_B  T / 6\pi\mu r,
\label{eq:einstein}
\end{align}
\noindent
where $k_B=1.381 \times10^{-23} \mathrm{m^2kgs^{-2}K^{-1}}$ is the Boltzmann constant, $T=298 K$ is the absolute colloidal temperature, and $\mu=1.1\times10^{-3}$  Pa.s is the viscosity of the liquid (a milliQ water-$\mathrm{D_2O}$ mixture 1:1). For the particles (radius, $r=2.5 \times 10^{-7}$) used in the experiment the theoretical prediction is $D=7.94\times10^{-7} \mathrm{mm^2/s}$. The experimentally measured diffusion coefficient $D=1.4\times10^{-7} \mathrm{mm^2}/s$ for the confined medium is 5.6 times smaller than the value predicted in equation \ref{eq:einstein} as expected, since the colloids cannot be considered suspended in the bulk due to the close proximity to no-slip boundary at the grain walls.
\section*{Trapping time distribution in a single vortex}

\begin{figure}[htb!]
    \centering
	\includegraphics[width=0.7\linewidth]{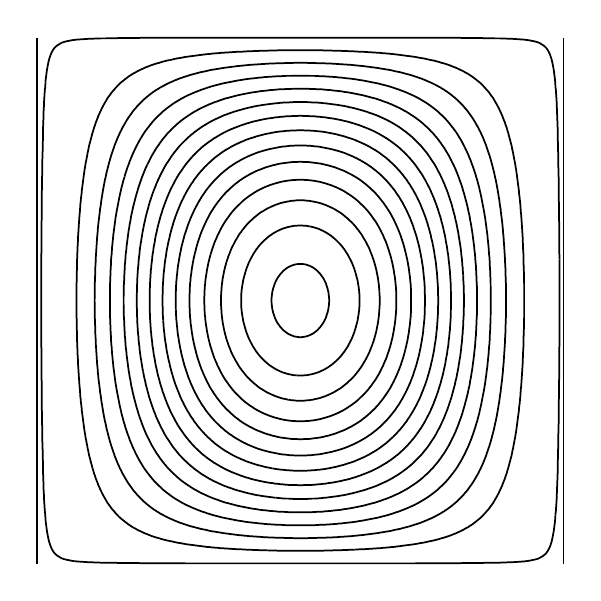}
	\caption{Streamlines corresponding to the streamfunction $\psi(\vec{x}) = 16 \sin(\pi y) x^2 (x - 1)^2$~\cite{young1989}, which is characterized by no-slip at the vertical boundaries at $x = 0, 1$. The spacing of the streamlines decreases with distance from the vertical boundaries, while it is approximately constant at the horizontal boundaries. \label{fig:roll}}
\end{figure}

We consider here the trapping times in a vortex that is confined at the vertical boundaries by solid walls and open at the horizontal boundaries, see Figure~\ref{fig:roll}. In our discussion, we follow the analyses of~\cite{young1989} and~\cite{Bouchaud1990}. To this end, we consider the advection-diffusion (mass conservation) equation for the solute concentration $c(\vec{x},t)$,

\begin{align}\label{ade}
	\frac{\partial c(\vec{x},t)}{\partial t} + \vec{v}(\vec{x}) \cdot \nabla c(\vec{x},t) - D \nabla^2 c(\vec{x},t) = 0, 
\end{align}
\noindent
for an instantaneous solute pulse $c(\vec{x},t = 0) = \delta(t)$ at the upper boundary of the roll. The solute advects and diffuses into the convection roll. The trapping time is given by the solute return time distribution from the vortex to the upper boundary. For pure diffusion, i.e., for $\vec{v}(\vec{x}) \equiv \vec{0}$, the trapping time distribution scales as $\phi(t) \sim t^{-3/2}$~\cite[][]{Redner2001}. The same behavior would be obtained if the flow in the vortex were simply a solid body rotation, see also below. Here this is different due to the shear flow along the solid boundaries. The distance between streamlines is larger in the shear flow close to the solid boundaries than at the horizontal boundaries, see Figure~\ref{fig:roll}. This means, the distance a particle diffuses inside the vortex across streamlines at the vertical boundary, is decreased at the horizontal boundary due to the compression of the streamlines. This interplay leads to a steeper decay of the trapping time distribution as $\phi(t) \sim t^{-5/3}$ because particles have the chance to leave the vortex earlier than in the absence of shear advection at the solid boundaries.

To show this quantitatively, we first note that during a rapid initial phase, solute is spread uniformly along and between streamlines due to shear dispersion~\cite[][]{rhines1983,young1989}. In order to formalize this  observation, we consider the streamline coordinates $(\psi,\phi)$, where $\psi(\vec{x})$ is the stream function, so that

\begin{align}
	\vec{v}(\vec{x}) = \left[-\frac{\partial}{\partial y}, \frac{\partial}{\partial x} \right]^\top \psi(\vec{x}),
\end{align}
\noindent
the superscript $\top$ denotes the transpose and $\nabla \psi(\vec{x}) \cdot \vec{v}(\vec{x}) = 0$. Furthermore, $\nabla \phi(\vec{x}) \cdot \vec{v}(\vec{x}) = v(\vec{x}) \equiv |\vec{v}(\vec{x})|$. Thus, the coordinate $\psi$ labels the streamline and $\phi$ denotes the distance along the streamline. We write $c(\vec{x},t) = c[\psi(\vec{x}),\phi(\vec{x},t)]$ and use for simplicity of notation the same letter for concentration in Cartesian and streamline coordinates. The concentration is constant along streamlines, this means that $c(\vec{x},t)$ can be represented by its average along a closed streamline as

\begin{align}\label{averagec}
	c(\vec{x},t) = \overline c(\psi,t) = \frac{1}{T(\psi)} \oint \frac{d \phi}{v(\psi,\phi)} c(\psi,\phi,t), && T(\psi) = \oint \frac{d \phi}{v(\psi,\phi)}.
\end{align}
\noindent
The integral is along the closed contour $\psi = \psi(\vec{x})$. Note that division by $v$ in the integrand accounts for the variable spacing between streamlines because the amount of solute between streamlines is constant. We now average the advection-diffusion equation~\eqref{ade} along a streamline and use~\eqref{averagec}, which gives

\begin{align}\label{adepsi}
	T(\psi) \frac{\partial \overline c(\psi,t)}{\partial t} - D \frac{\partial}{\partial \psi} \int d \vec{x} \nabla^2 \overline c[\psi(\vec{x}),t] H[\psi - \psi(\vec{x})] = 0,  
\end{align}
\noindent
where we used the identity

\begin{align}
	\frac{\partial}{\partial \psi} \int d \vec{x} F(\vec{x}) H[\psi - \psi(\vec{x})] = \int d \vec{x} F(\vec{x}) \delta[\psi - \psi(\vec{x})] = \oint \frac{d \phi}{v(\psi,\phi)} F(\psi,\phi). 
\end{align}
\noindent
where we used the fact that $v = |\nabla \psi|$. Using the Gauss theorem, the integral term on the right side of
Eq.~\eqref{adepsi} can be written as

\begin{align}
	\int d \vec{x} \nabla^2 \overline c[\psi(\vec{x}),t] H[\psi - \psi(\vec{x})] = \oint d \phi \vec{n} \cdot \nabla \overline c[\psi(\vec{x}),t] = \oint d \phi \vec{n} \cdot \nabla \psi(\vec{x}) \frac{\partial}{\partial \psi} \overline c(\psi,t).
\end{align}
\noindent
Thus, Eq.~\eqref{adepsi} becomes

\begin{align}\label{adepsi:2}
	T(\psi) \frac{\partial \overline c(\psi,t)}{\partial t} - D \frac{\partial}{\partial \psi} \left[C(\psi)\frac{\partial}{\partial \psi} \overline c(\psi,t) \right] = 0,  
\end{align}
\noindent
where we defined

\begin{align}
	C(\psi) = \oint d \phi v(\psi,\phi). 
\end{align}
\noindent
In order to continue we need to specify $T(\psi)$, which is the advection time along a closed contour $\psi$ and $C(\psi)$, the circulation of the velocity field. If both quantities where constant,
Eq.~\eqref{adepsi} would describe a simple diffusion equation and the trapping time distribution would scale as $\phi(t) \sim t^{-3/2}$. Here this is different. We approximate the two quantities
by noting that the main contributions to $T(\psi)$ come from the vertical sections close to the no-slip boundary. There, we approximate the streamfunction by $\psi(x) = \sigma x^2/2 $, where $\sigma$ is the shear rate. This implies that $v(\psi) \sim \sqrt{\psi}$ and $T(\psi) \sim \psi^{-1/2}$. The circulation $C(\psi)$ is dominated by sections along the horizontal boundaries, close to which the velocity field is approximately constant. This means $C(\psi) \approx C_0$. Equation~\eqref{adepsi:2} can then be written as

\begin{align}\label{adepsi:3}
	T(\psi) \frac{\partial \overline c(\psi,t)}{\partial t} - D_0 \frac{\partial^2}{\partial \psi^2} \overline c(\psi,t) = 0,  
\end{align}
\noindent
where $D_0 = D C_0$. We note that~\eqref{adepsi:3} is equivalent to the Langevin equation

\begin{align}
	d \psi = \sqrt{D_0 \frac{dt}{T(\psi)}} \xi(t),
\end{align}
\noindent
where $\xi(t)$ is a Gaussian white noise. We define the operational time $ds = dt/T(\psi)$ and write 

\begin{align}
	d \psi = \sqrt{D_0 ds} \xi(s), && dt = T(\psi) ds.
\end{align}
\noindent
Thus, the time $t(s)$ is given by

\begin{align}
	t(s) = \int\limits_0^s ds' T[\psi(s')] \approx \int\limits_0^s ds' \langle T[\psi(s')] \rangle,
\end{align}
\noindent
where the angular brackets denote the noise average. As $\psi(s)$ describes a Brownian motion, we obtain the scaling

\begin{align}
	\langle T[\psi(s)] \rangle \sim \int\limits_0^\infty d\psi \psi^{-1/2} \frac{\exp[-\psi^2/(4 D_0 s)]}{\sqrt{4\pi D_0 s}} \sim s^{-1/4},
\end{align}
\noindent
which implies that 

\begin{align}\label{ts}
	t(s) \sim s^{3/4},
\end{align}
\noindent
i.e., time increases slower than linearly with operational time.

\subsection*{Trapping times}

We want to determine the distribution of trapping times, which implies the distribution of $t(s)$ such that $s$ is the operational time to escape the roll, this means to return to the streamline $\psi = 0$ after entering the role at $s = 0$. This means we need to determine the return time PDF. As this problem is not well-defined in continous space, we consider instead the first passage time distribution from a position $\psi$ to the edge. For $\psi \ll 1$, this approximates the return time distribution. Since $\psi(s)$ describes a Brownian motion, the first passage time distribution from a position $\psi$ to the edge of the role can, for times small comapred to the diffusion time across the roll, be approximated by the inverse Gaussian

\begin{align}\label{gig}
	g(s,\psi) = \frac{\psi \exp\left(-\frac{\psi^2}{4 D s}\right)}{\sqrt{2 \pi D s^3}}.   
\end{align}
\noindent
This means, it scales as $g \sim s^{-3/2}$. The trapping time distribution $\phi(t)$ is obtained from $g(s)$ by variable transform

\begin{align}
	\phi(t) dt = g[s(t)] \frac{ds(t)}{dt} dt.  
\end{align}
\noindent
This implies for $s(t) \sim t^{4/3}$ and $g(s) \sim s^{-3/2}$ the scaling

\begin{align}
	\phi(t) \sim t^{-5/3}. 
\end{align}
\noindent
This behavior is valid as long as the solute is distributed along the outer streamlines of the vortex. For times larger than the diffusion time $\tau_D = R^2/D$ across the extension $R$ of the roll, $\phi(t)$ is decaying exponentially fast.

\subsection*{Residence times}

In order to obtain the residence time distribution, we consider a uniform distribution of particles inside the role. Note that a uniform distribution $p_0(x) = 1$ in equidistant space $x$ implies in streamline coordinates that

\begin{align}
p_0(\psi) \sim \psi^{-1/2}. 
\end{align}
\noindent
Thus, we obtain the residence time distribution by integration of~\eqref{gig} because it describes the residence time distribution of particles that originate from $\psi$. Thus, we obtain 

\begin{align}\label{Gs}
G(s) = \int d\psi \psi^{-1/2} \frac{\psi \exp\left(-\frac{\psi^2}{4 D s}\right)}{\sqrt{2
		\pi D s^3}} = \int d\psi \psi^{1/2} \frac{\exp\left(-\frac{\psi^2}{4 D s}\right)}{\sqrt{2
		\pi D s^3}} \sim s^{-3/4}. 
\end{align}
\noindent
Thus, the residence time distribution behaves as

\begin{align}\label{Phi}
	\Phi(t) dt = G[s(t)] \frac{ds(t)}{dt} dt.  
\end{align}
\noindent
The scaling of $\Phi$ is obtained substituting eq.~\eqref{ts} and eq.~\eqref{Gs} into eq.~\eqref{Phi}, leading to

\begin{align}
	\Phi(t) \sim t^{-2/3}. 
\end{align}
\noindent
This is the scaling observed for the residence time from the experimental and numerical data.

\section*{Continuous time random walk model}
\label{sc:ctrw}

The breakthrough curves are fully determined by the particle retentention at the first step. Thus, we model the breakthrough curves through the CTRW

\begin{align}\label{ctrw}
	x_{n+1} = x_n + \ell_c + \sqrt{2 D^\ast \tau_n} \xi_n, && t_{n+1} = t_n + \tau_n. 
\end{align}
\noindent
With the distribution $p_0(x,t)$ of initial particle positions and times $(x_0,t_0)$ given by 

\begin{align}
	p_0(x,t) = \frac{1}{L} \mathbb I(0 < x < L) \left[(1 - \alpha) \delta (t) + \alpha \Phi(t) \right]. 
\end{align}
\noindent
where $\alpha$ is the proportion of particles initially located in the DEPs, and $\mathbb I(\cdot)$ is an impulse function, which is $1$ if its argument is true and $0$ else. This means that the initial positions $x_0$ are uniformly distributed in space, and the residence time within a DEP at the initial position $x_0$ is given by a random time $t_0$. Its distribution is given by $\Phi(t)$, which is the residence time within a DEP of depth $\Lambda \lambda_m$ and so it is the $\Gamma$ distribution observed with our direct numerical simulation

\begin{equation}
	\Phi(t) = \frac{(t/\tau_D)^{-2/3} \exp(-t/\tau_D)}{\Gamma(1/3)}.
\end{equation}
\noindent
whose -2/3 scaling was we determined in the previous section. The time increment $\tau_n$ is set constant and equal to the advection time over the distance $\ell_c$ as $\tau_n = \Delta t = \ell_c/ u_m$ with $u_m$ the average velocity in percolating pores. The random displacement $\xi_n$ is a Gaussian random variable with 0 mean and unit variance, and $D^\ast$ is the fitted dispersion coefficient of the transmitting pore network. \\

In order to determine the breakthrough curve at all times for this CTRW model, we first consider the BTC for particles that originate at a distance $x$ from the outlet. If the particle is not in a
DEP, the time to reach the outlet is distributed according to the inverse Gaussian distribution~\cite{Dentz2011}

\begin{align}\label{ig}
	f_0(t,x) = \frac{x \exp\left[-\frac{(x - v_m t)^2}{4 D^\ast t}\right]}{\sqrt{2\pi D^\ast t^3}}
\end{align}
\noindent
because transport in this CTRW model is given by advection and dispersion. If the particle is initially trapped, the time to arrive at the outlet is $t_0 + t_{f}$, where $t_{f}$ is distributed according to the inverse Gaussian~\eqref{ig} because transport is due to advection and dispersion once the particle has been released from the DEP. Thus, the BTC for particle originating at a distance $x$ from the outlet can be written as

\begin{align}
	f(t,x) = (1 - \alpha) f_0(t,x) + \alpha \int\limits_0^t dt' f_0(t - t',x) \Phi(t'). 
\end{align}
\noindent
The full breakthrough curve $F(t)$ is obtained by integration of $f(t,x)$ over the full range $L$ such that

\begin{align}
	F(t) = \frac{1}{L} \int\limits_0^L dx f(t,x). 
\end{align}
\noindent
In order to extract the tailing behavior, we note that $f(t,x)$ is peaked about $t = x/v$ so that we can write

\begin{align}
	f(t,x) = (1 - \alpha) f_0(t,x) + \alpha \Phi(t - x/v)
\end{align}
\noindent
Thus, we can write for the breakthrough curve

\begin{align}\label{analytical_btc}
	F(t) = (1 - \alpha) \frac{1}{L} \int\limits_0^L dx f_0(t,x) + \alpha\frac{1}{L} \int\limits_{0}^{L} dx \Phi(t-x/v) 
\end{align}
\noindent
For times $t \gg L/v$, we obtain

\begin{align}
	F(t) = (1 - \alpha) \frac{1}{L} \int\limits_0^L dx f_0(t,x) + \alpha \Phi(t) 
\end{align}
\noindent

\section*{Initial colloidal distribution}

\begin{figure}[htb!]
\centering
	\includegraphics[width=0.8\linewidth]{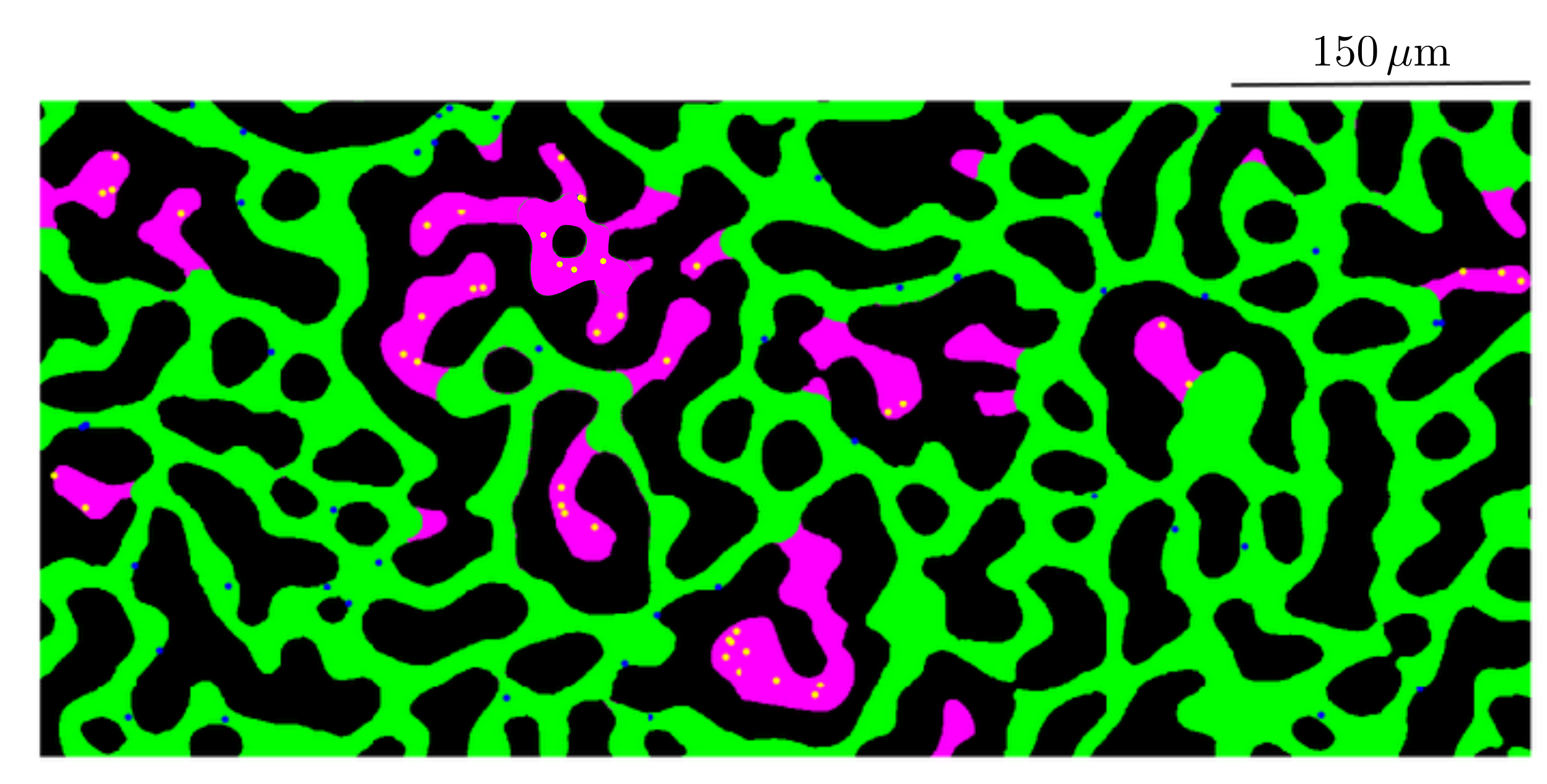}
	\caption{\textbf{Relative distribution of suspended particles between DEP and FC in experiment:} Location of suspended particles inside the dead-end pores (yellow dots in magenta regions) and transmitting pores (blue dots in green regions) in a subsection of the porous medium. The fraction of total particle counts is $\alpha$ = 0.22 in DEP and $1-\alpha = 0.78$ in TP.  \label{sfig7}}
\end{figure}

The concentration of colloids after the saturation phase, that lasted 24 hours, resulted non-homogeneously distributed across the medium.  To quantify this distribution in terms of $\alpha$, we consider a representative section (15 mm $\times$ 4 mm) from the experiment and measure the number of suspended (mobile) colloids using the method described in Methods $e$.  Then, using the corresponding segregation map (see Method $a$), we compute the fraction of colloids located in DEP areas (see Figure \ref{sfig7}): the TPs have a slightly smaller colloids concentration with respect to the DEPs, resulting in 22\% of the total suspended colloids accumulated in the DEPs and 78\% of the colloids in the TPs (while DEPs represent the 6\% of the total volume of the porous system and the TPs represent the 94\% of it). This means that the fraction of colloids initially located within a DEP is $\alpha = 0.22$ and, as a consequence, $1-\alpha$ is the fraction of colloids initially located within a TP. We took this $\alpha$ value into consideration to implement the CTRW model for colloids transport and the  BTC as discussed in section \ref{sc:ctrw}. \\

\subsection*{References}
\bibliographystyle{unsrt}
\bibliography{library_SI}




\begin{figure}[htb!]
	\centering
	\includegraphics[width=1\linewidth]{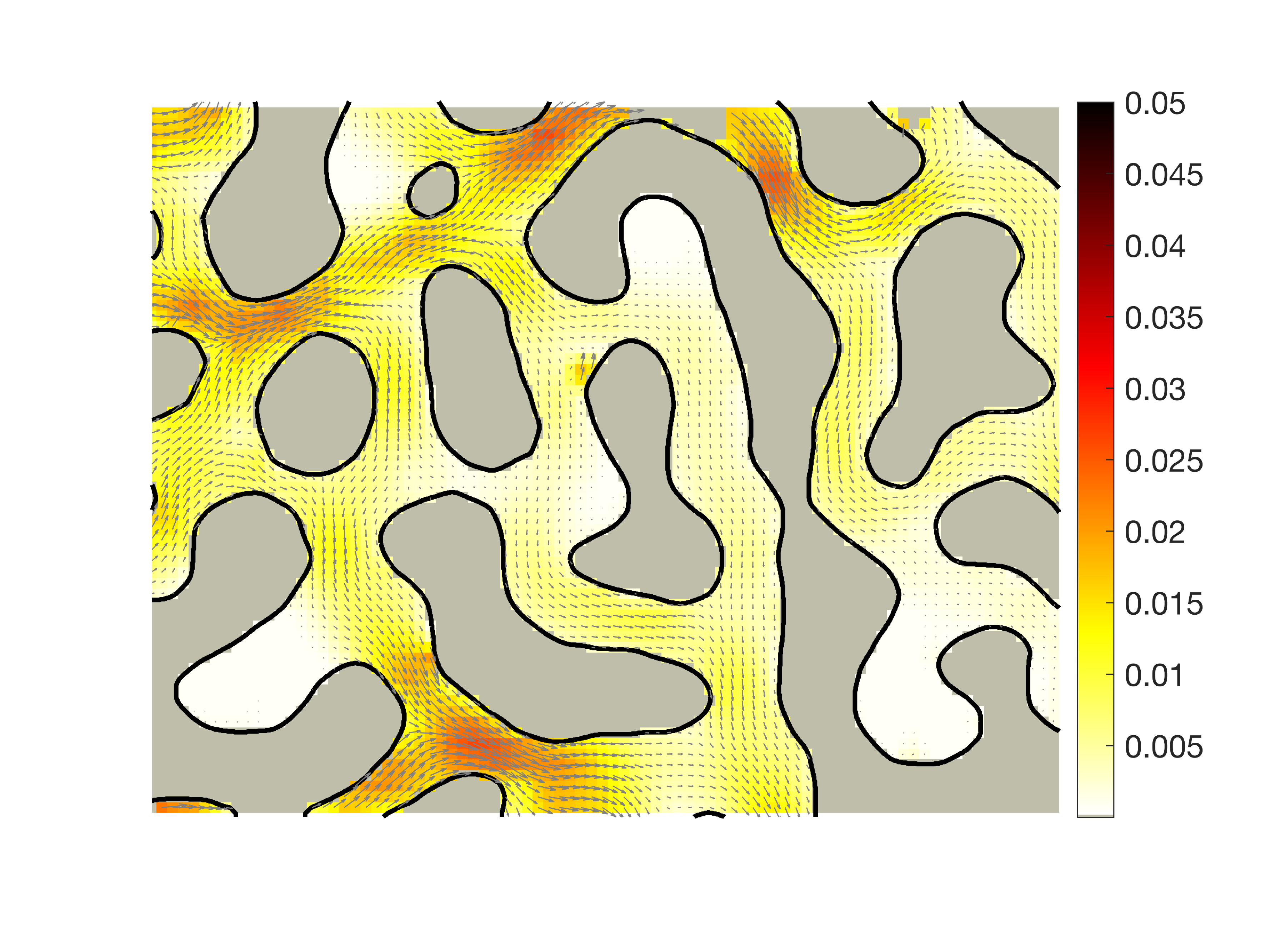}
	\caption{\textbf{Velocity field measured via particle image velocimetry (PIV):} Absolute value of the velocity field (mm/s) superposed with velocity vectors of a portion of our system. The PIV experiment is performed following \cite{deAnnaNaturePhys2021}, seeding with 1 $\mu$m beads in water flowing at overall flow rate $Q = 0.5 \mu L/min$.}
	\label{sfig1}
\end{figure}

\begin{figure}[htb!]
	\centering
	\includegraphics[width=1\linewidth]{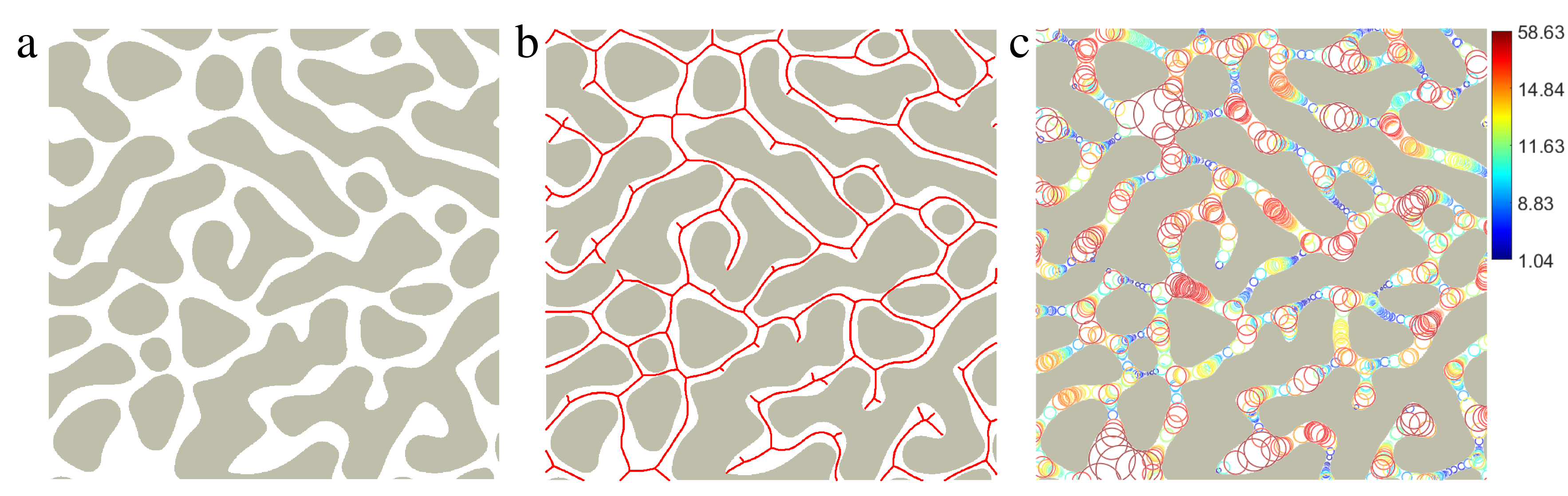}
	\caption{\textbf{Method of maximum inscribed circle (MIC) to compute the pore-size distribution across the porous medium:} \textbf{a,} binary image of  a small sub-section of the model porous medium with white and gray colors representing the pore-space and the constitutive grains, respectively, \textbf{b,} pixel location of the skeletonized structure (red lines) of the pore-space region, and \textbf{c,} circles fitted within the pore-space region using maximum inscribed circle method~\cite{Birdal2021} and color-coded with their radii ($\mu$m).}
	\label{sfig2}
\end{figure}

\begin{figure}[htb!]
	\centering
	\includegraphics[width=1\linewidth]{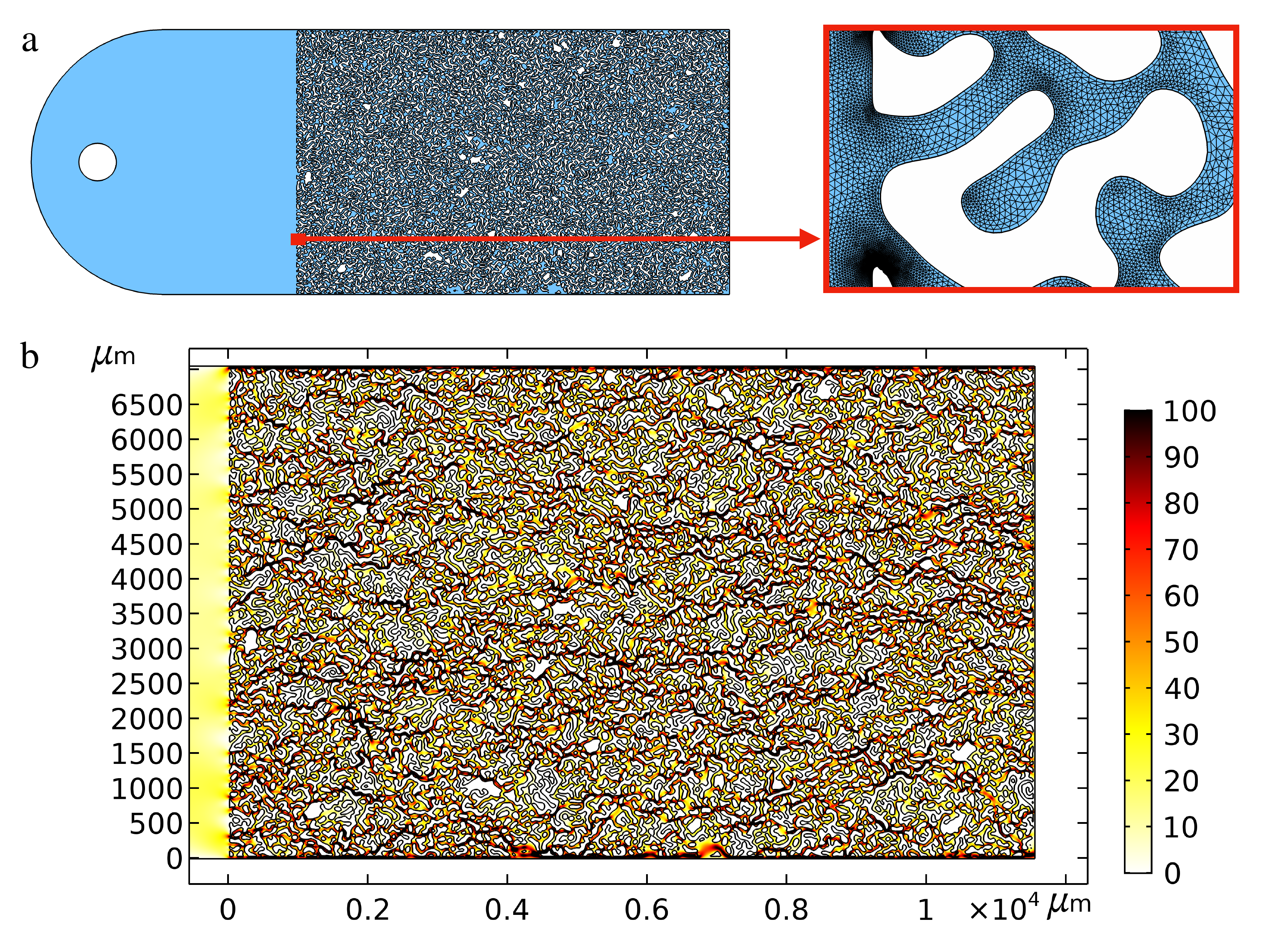}
	\caption{\textbf{Eulerian velocity field numerical solution of the Stokes flow equations:} \textbf{a,} Computation domain that contains a 2D porous medium of length 11 mm and width 7 mm. Inset shows an enlarged subsection of the discretized domain having an adaptive tetrahedral mesh with minimum exlement size of 0.14 microns near each rigid surface, such that the boundary layers are well resolved. \textbf{b,} Magnitude of computed Eulerian velocity ($\mu\mathrm{m/s}$). }
	\label{sfig3}
\end{figure}

\begin{figure}[htb!]
	\centering
	\includegraphics[width=1\linewidth]{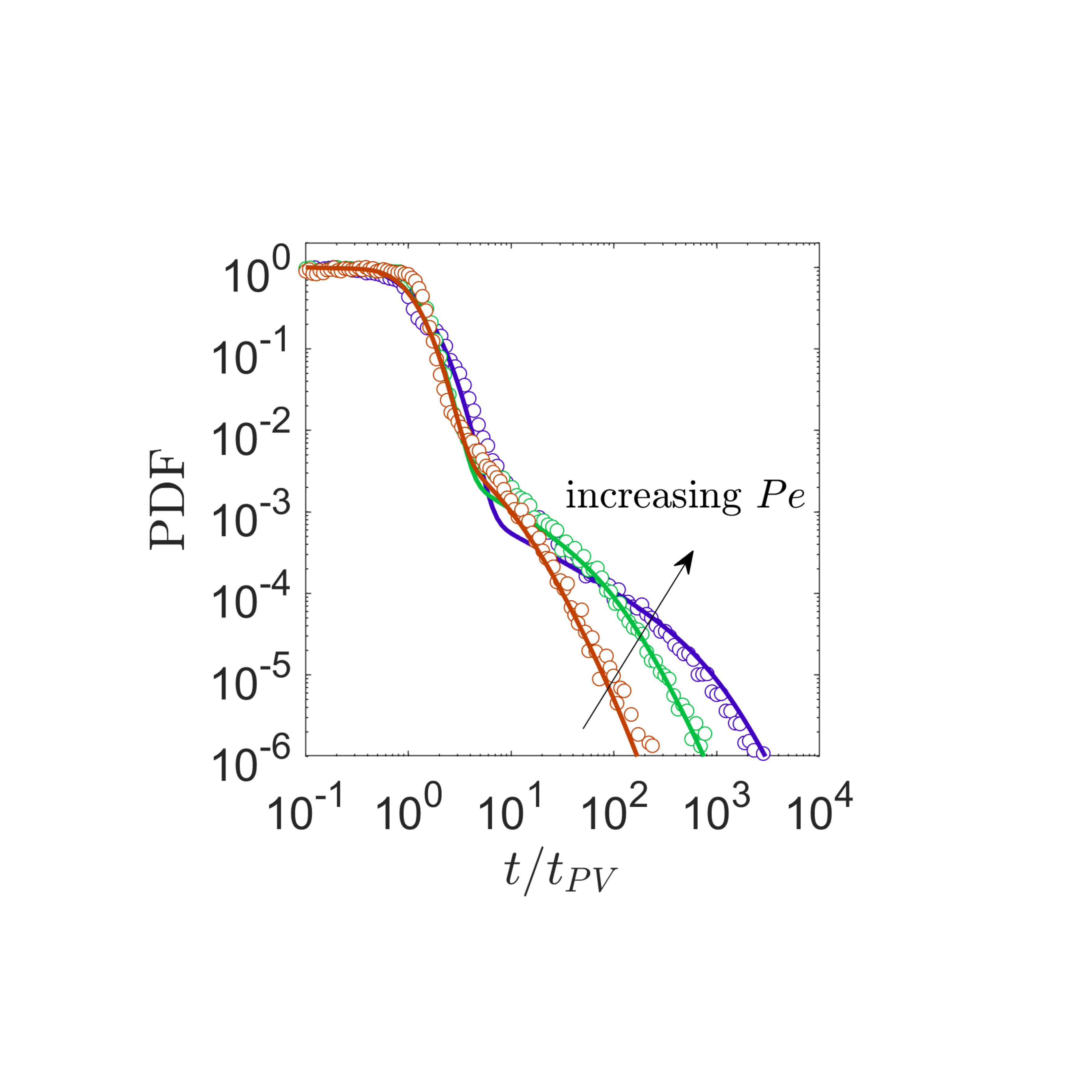}
	\caption{\textbf{Effect of P\'eclet number on BTC}: Probability density function (PDF) of particle escape time (equivalent to the BTC) versus normalized time ($t/t_{PV}$) for P\'ectlet number, $Pe$ = 68, 680 and 6800 obtained from tracking $\mathrm{10^5}$ homogeneously distributed particles in the simulated velocity field (symbols) and the analytical  (CTRW) model (solid line). The P\'eclet number is varied by changing $D$ across three orders of magnitude.}
	\label{sfig4}
\end{figure}

\begin{figure}[htb!]
	\centering
	\includegraphics[width=1\linewidth]{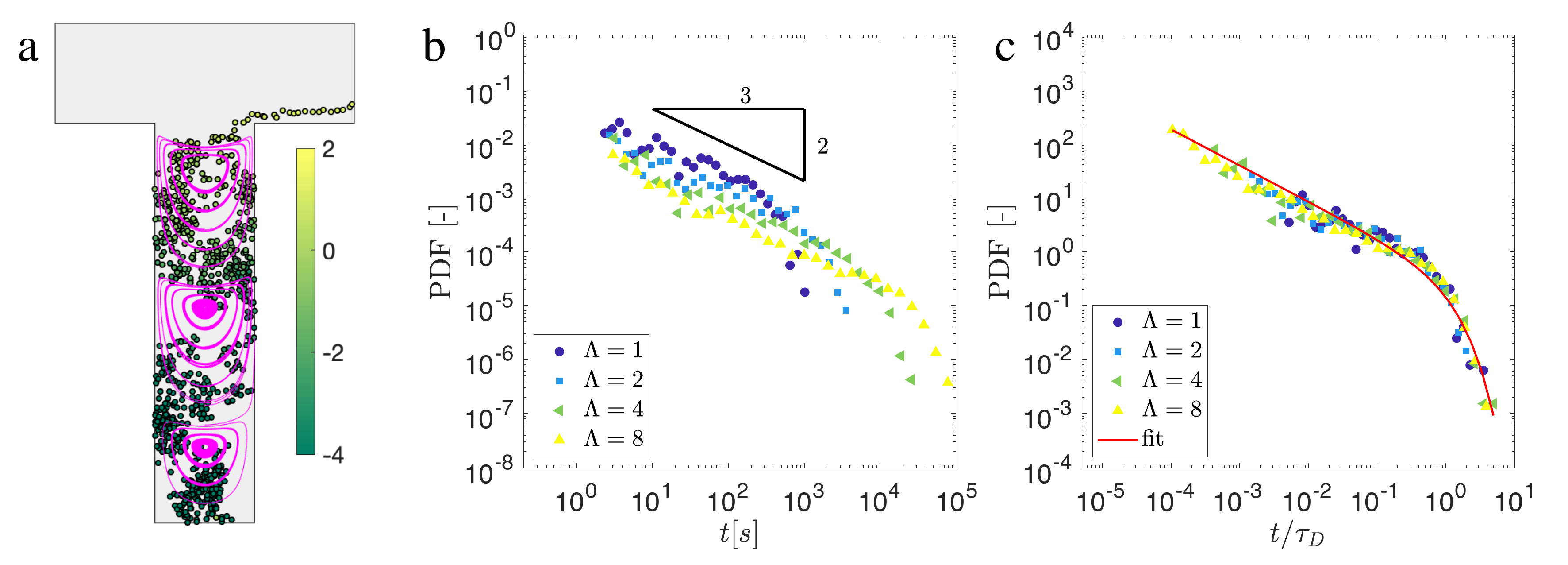}
	\caption{\textbf{Simplified 2D model that captures the particle escape time from the dead-end pores}: \textbf{a,} Numerically simulated trajectory (color-coded with local Peclet number $Pe^* = \lambda_mv_p/D_m$) of a particle escaping from the bottom of a  2D rectangular cavity (aspect ratio, $\Lambda$ = 4) connected to a  channel.  A series of closed streamlines highlights the vortex flow structure inside the cavity, and \textbf{b,} probability density function (PDF) of escape time of particles of a single cavity for $\Lambda = 1, 2, 4, 8$, and \textbf{c,} the same as \textbf{b} with time rescaled with diffusion time-scale $\tau_D = (\lambda_m\Lambda)^2/D$.}
	\label{sfig4}
\end{figure}

\begin{figure}[htb!]
\centering
	\includegraphics[angle = 0,width=0.5\linewidth]{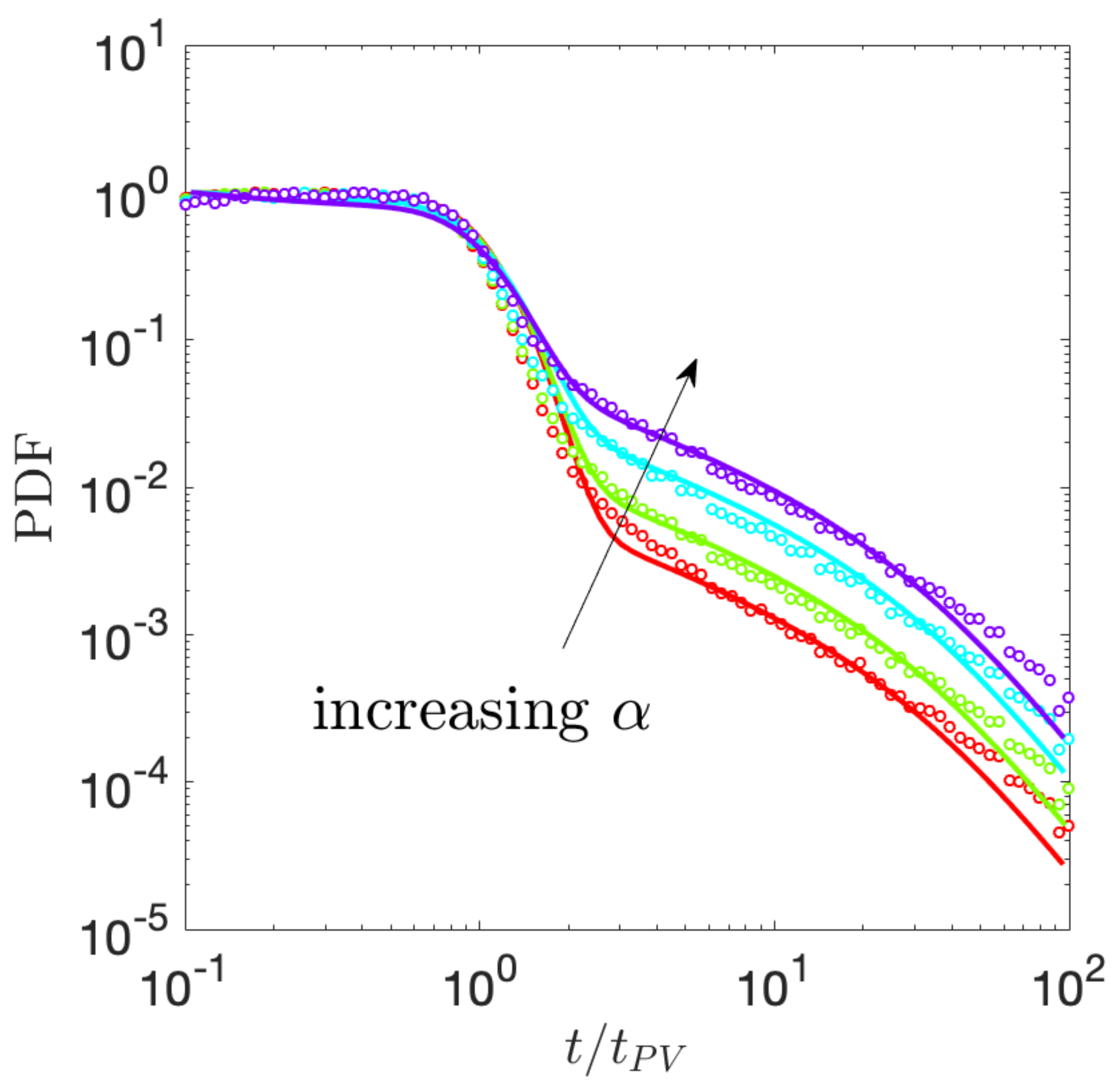}
	\caption{\textbf{The initial abundancy $\alpha$  of suspended particles between DEP and TP influences the overall BTC:} Probability density functions (PDF) of particle arrival time in hyperuniform porous medium based on numerical simulation (circles) and the analytical model in equation \ref{analytical_btc} (lines) for four values of $\alpha$ = [0.06, 0.11, 0.24, 0.38].   \label{sfig8}}
\end{figure}